\pdfoutput=1
\documentclass[aps,prl,reprint,superscriptaddress]{revtex4-1}

\usepackage[dvipsnames]{xcolor}
\usepackage{delimset}
\usepackage{accents}
\usepackage{dsfont}
\usepackage{scalerel}
\usepackage{tikz}
\usepackage{delimset}

\usepackage{amsmath,amssymb}

\newcommand{\pd}{\partial}
\newcommand{\mE}{\mathcal{E}}
\newcommand{\mA}{\mathcal{A}}

\newcommand{\OO}{\mathcal{O}}

\newcommand{\hOmega}{\omega}
\newcommand{\tOmega}{\widetilde{\Omega}}

\newcommand{\RR}{\mathbb{R}}

\newcommand{\inv}{^{-1}}

\newcommand{\order}[1]{\mathcal{O}\brk{#1}}

\newcommand{\ost}[2]{\accentset{\scriptscriptstyle{\scaleto{\brk{\hspace{-0.06em}{\mathrm{#2}}\hspace{-0.06em}}}{4.5pt}}}{#1}{}} 

\newcommand{\group}[2]{\MakeUppercase{#1}\brk{#2}}
\newcommand{\algebra}[2]{\MakeLowercase{#1}\brk{#2}}

\newcommand{\der}{d}
\newcommand{\D}{D}

\newcommand{\gOmega}{\Omega}

\newcommand{\gQ}{\ost{Q}{G}}

\newcommand{\nbi}{%
  \affiliation{%
    The Niels Bohr Institute,
    University of Copenhagen,
    Blegdamsvej 17,
    DK-2100 Copenhagen Ø,
    Denmark
}}
\newcommand{\zur}{%
  \affiliation{%
    Institut für Theoretische Physik,
    ETH Zürich,
    Wolfgang-Pauli-Strasse 27, 8093 Zürich, Switzerland
}}
\newcommand{\edi}{%
  \affiliation{%
    School of Mathematics and Maxwell Institute for Mathematical Sciences,
    University of Edinburgh,
    Peter Guthrie Tait Road,
    Edinburgh EH9 3FD, UK
}}
\newcommand{\nor}{%
  \affiliation{%
    Nordita,
    KTH Royal Institute of Technology and Stockholm University,
    Roslagstullsbacken 23,
    SE-106 91 Stockholm,
    Sweden
}}

\usepackage{hyperref}

\begin{document}
\title{%
  Galilean first-order formulation for the non-relativistic expansion of general relativity
}
\author{Dennis Hansen}\zur
\author{Jelle Hartong}\edi
\author{Niels A. Obers}\nbi\nor
\author{Gerben Oling}\nbi


\begin{abstract}
  We reformulate the Palatini action for general relativity (GR) in terms of moving frames that exhibit local Galilean covariance in a large speed of light expansion.
  For this, we
  express the action
  in terms of variables that are adapted to a Galilean subgroup of the $\group{GL}{n,\RR}$ structure group of a general frame bundle.
  This leads to a novel Palatini-type formulation of GR
  that provides a natural starting point for a first-order non-relativistic expansion.
  In doing so, we show how a comparison of Lorentzian and Newton--Cartan metric-compatibility explains the appearance of torsion in the non-relativistic expansion.
\end{abstract}

\maketitle

\paragraph{Introduction.}
In recent years the study of non-relativistic approximations to general relativity (GR) has gained renewed interest.
As an important development, a geometric description of the non-relativistic expansion of general relativity in inverse powers of the speed of light~$c$ was obtained in Refs.~\cite{VandenBleeken:2017rij,Hansen:2019pkl,Hansen:2020pqs}, building on earlier work~\cite{Dautcourt:1996pm,Tichy:2011te}.
To a large extent, this progress was made possible by new insights \cite{Andringa:2010it,Christensen:2013lma,Hartong:2015zia} in \emph{Newton--Cartan} geometry \cite{Cartan1,Cartan2}, which
takes the place of the Lorentzian geometry of GR at leading order in the non-relativistic expansion.
The resulting geometric non-relativistic expansion of gravity can be carried out on the level of the action, which means that it can be applied to any astrophysical setting where velocities are small.
In particular, the $1/c^2$ expansion of GR holds the prospect of being related to a covariant and off-shell formulation of the post-Newtonian expansion, since the leading term has been shown to reproduce the 1PN approximation~\cite{Tichy:2011te}.
More generally, covariant theories of gravity involving Newton--Cartan-type geometries prominently appear
in recent developments in non-relativistic field theory~\cite{Son:2013rqa,Jensen:2014aia,Hartong:2014oma,Geracie:2014nka},
non-relativistic string theory and limits of the AdS/CFT correspondence~\cite{Andringa:2012uz,Harmark:2017rpg,Bergshoeff:2018yvt}, along
with many other areas that exhibit non-relativistic physics.

In this work, we focus on the $1/c^2$ expansion of GR, which has been shown to lead to a modification of the original notion of Newton--Cartan (NC) geometry, known as type~II torsional Newton--Cartan (TNC) geometry~\cite{Hansen:2019pkl,Hansen:2020pqs},
as the correct framework for a covariant action of non-relativistic gravity.
When coupled to a point particle, the equations of motion of this action lead to the Poisson equation of Newtonian gravity in an arbitrary frame.
In addition, for appropriate matter sources, the theory generalizes Newtonian gravity since it includes the effects of gravitational time dilation due to strong gravitational fields~\cite{VandenBleeken:2017rij,Hansen:2020pqs,Hansen:2019vqf}.
When time is absolute, type~II TNC geometry reproduces the usual NC geometry.

To set the stage, we briefly review the non-relativistic expansion of GR in powers of $1/c^2$ following Ref.~\cite{Hansen:2020pqs}.
We can make the factors of $c$ in the metric explicit by introducing the \emph{`pre-non-relativistic'} (PNR) parametrization
\begin{equation}
  \label{eq:lorentzian-tangent-metric-pnr-parametrization}
  g_{\mu\nu}
  = - c^2 T_\mu T_\nu + \Pi_{\mu\nu},
  \quad
  g^{\mu\nu}
  = - \frac{1}{c^2} V^\mu V^\nu + \Pi^{\mu\nu}.
\end{equation}
The resulting PNR variables $(T_\mu, \Pi_{\mu\nu})$ and $(V^\mu, \Pi^{\mu\nu})$ are all of order $\order{c^0}$.
They satisfy the orthogonality and completeness relations
\begin{subequations}
  \label{eq:orthogonality-completeness-relations-metric}
  \begin{gather}
    V^\mu T_\mu = -1,
    \quad
    T_\mu \Pi^{\mu\nu} = 0,
    \quad
    V^\mu \Pi_{\mu\nu} = 0,
    \\
    \delta^\mu_\nu = - V^\mu T_\nu + \Pi^{\mu\rho} \Pi_{\rho\nu},
  \end{gather}
\end{subequations}
which imply in particular that $\Pi^{\mu\nu}$ has rank $n-1$, where $n$ is the spacetime dimension.
Assuming analyticity, they can be expanded in powers of~$1/c^2$, which leads to
\begin{subequations}
  \label{eq:pnr-metric-expansions}
  \begin{align}
    T_\mu
    &= \tau_\mu + \frac{1}{c^2} m_\mu + \OO(1/c^4),
    \\
    \Pi_{\mu\nu}
    &= h_{\mu\nu} + \frac{1}{c^2} \Phi_{\mu\nu} + \OO(1/c^4).
  \end{align}
\end{subequations}
Likewise, $V^\mu=v^\mu + \OO(c^{-2})$ and $\Pi^{\mu\nu} = h^{\mu\nu}+\OO(c^{-2})$.
The PNR variables transform under local Lorentz transformations, which reduce to Galilean transformations in the $1/c^2$ expansion.
Under these Galilean transformations, $(\tau_\mu,h^{\mu\nu})$ are invariant, while $(v^\mu,h_{\mu\nu})$ transform.

Before expanding in powers of~$1/c^2$, the pre-non-relativistic variables simply provide a parametrization of the Lorentzian metric through Equation~\eqref{eq:lorentzian-tangent-metric-pnr-parametrization}.
Upon expanding, however, the resulting leading-order fields $\tau_\mu$ and $h^{\mu\nu}$
define a {\it NC geometry},
which is modified into a {\it type~II TNC geometry} by adding the subleading fields $m_\mu$ and $\Phi_{\mu \nu}$.
As such, the PNR parametrization~\eqref{eq:lorentzian-tangent-metric-pnr-parametrization} is a convenient way of recasting the Lorentzian metric variables of GR in such a way that the appropriate non-relativistic geometry appears naturally in the expansion.

Next, we rewrite the Einstein--Hilbert action of GR,
\begin{equation}
  \label{eq:standard-einstein-hilbert-action-lagrangian}
  S_\text{EH} = \frac{c^3}{16\pi G_N}  \int_{M} \mathcal{L}'_\text{EH}
  \sqrt{-g}\, \der^n x,
  \quad
  \mathcal{L}'_\text{EH} = R,
\end{equation}
in terms of the PNR variables~\eqref{eq:lorentzian-tangent-metric-pnr-parametrization}.
Here, $R$~is the Ricci scalar of the Levi-Civita connection, which is the unique torsion-free connection that preserves the Lorentzian metric $g_{\mu\nu}$ under parallel transport, as is convenient for describing the degrees of freedom of GR.
However, to obtain a connection that is compatible with the non-relativistic geometry arising from the expansion
\eqref{eq:pnr-metric-expansions}, it is convenient to instead use the PNR connection
\begin{equation}
  \label{eq:C-connection}
    C^{\rho}_{\mu\nu}
    = -V^\rho\partial_\mu T_\nu
    +\Pi^{\rho\sigma}\partial_{(\mu}\Pi_{\nu)\sigma}
    - \frac{1}{2} \Pi^{\rho\sigma}\partial_\sigma\Pi_{\mu\nu}\,,
\end{equation}
which leads to a covariant derivative under which $T_\mu$ and $\Pi^{\mu\nu}$ are covariantly constant.
This has the consequence that the Newton--Cartan structure $(\tau_\mu, h^{\mu\nu})$ arising from these variables at leading order in the $1/c^2$ expansion is covariantly constant with respect to
the covariant derivative coming from the leading-order term in the expansion of $C^\rho_{\mu\nu}$.
Note that the connection \eqref{eq:C-connection} has nonzero torsion.

Up to a total derivative, the Einstein--Hilbert Lagrangian then takes the form
\begin{equation}
  \label{eq:pnr-einstein-hilbert-lagrangian}
  \mathcal{L}'_{\rm EH}
  = \frac{c^2 }{4}\Pi^{\mu\nu}\Pi^{\rho\sigma} T_{\mu\rho} T_{\nu\sigma}
  +\Pi^{\mu\nu}\ost{R}{C}_{\mu\nu}
  - \frac{1}{c^2} V^\mu V^\nu\ost{R}{C}_{\mu\nu},
\end{equation}
where $T_{\mu\nu}=(\der T)_{\mu\nu} = \partial_\mu T_\nu - \partial_\nu T_\mu$ and $\ost{R}{C}_{\mu\nu}$ is the Ricci tensor corresponding to the connection \eqref{eq:C-connection}.
Note that  $\sqrt{-g}$ is of order $c$, so the leading term in the action \eqref{eq:standard-einstein-hilbert-action-lagrangian} is ${\cal{O}} (c^6)$.
The parametrization~\eqref{eq:pnr-einstein-hilbert-lagrangian} is the starting point for the large speed of light expansion performed in Refs.~\cite{Hansen:2019pkl,Hansen:2020pqs}.
So far, the resulting expansion of the action has been carried out up to second subleading order,
\begin{equation}
  \mathcal{L}'_\text{EH}
  = c^2 \mathcal{L}'_\text{LO}
  + \mathcal{L}'_\text{NLO}
  + \frac{1}{c^2} \mathcal{L}'_\text{NNLO}
  + \OO(1/c^4).
\end{equation}
The leading-order term in the expansion, coming from the first term in Equation~\eqref{eq:pnr-einstein-hilbert-lagrangian}, imposes a particular constraint on the torsion of the geometry, as we will see below.
The NNLO equations of motion (which include the equations of motion of the preceding orders~\cite{Hansen:2019svu}) describe a type II torsional Newton--Cartan theory of gravity.
This theory reproduces Newtonian gravity and generalizes it to include strong gravity.
However, carrying out the expansion to higher orders becomes increasingly cumbersome in a second-order formulation.

In this Letter, our aim is to introduce the pre-non-relativistic form of the \emph{first-order} Palatini action of GR, which simplifies the computation of the $1/c^2$ expansion at higher orders.
In addition,
it is expected to simplify the study of a wide range of physical applications, for example by making the construction of boundary charges more accessible (see for example~\cite{Jacobson:2015uqa}), enabling the coupling to fermion fields, and simplifying Kaluza--Klein reductions.

Furthermore, we provide a new perspective on the geometric interpretation of the PNR parametrization using \emph{moving frames}, which clarifies the appearance of torsion and the local Galilean symmetries associated to Newton--Cartan geometry.
By embedding both the local Lorentz symmetry of GR and the local Galilean symmetry of NC~geometry inside the $\group{GL}{n,\RR}$ frame bundle and its associated general (linear) affine connections, we can translate between the corresponding notions of torsion and metric-compatibility.
This affine perspective also naturally connects to the `triality' between the usual metric formulation of GR and its equivalent formulations in terms of torsion or non-metricity~\cite{Hehl:1994ue,Jimenez:2019woj}.

\paragraph{Affine and Lorentzian connections.}
In the following, we will frequently use moving frames, which are also commonly known as vielbeine, to describe the geometry of the tangent bundle.
For a spacetime $M$, we denote by
\begin{equation}
  E^A = E^A_\mu \der x^\mu,
  \quad
  \Theta_A = \Theta_A^\mu \pd_\mu,
  \quad
  E^A(\Theta_B) = \delta^A_B,
\end{equation}
a set of vielbeine and their duals, where $A=0,\ldots,n-1$ are tangent spacetime indices.
Additionally, we introduce a connection $\Omega^A{}_B=\Omega_\mu{}^A{}_B \der x^\mu$ and its associated covariant derivative $\D$, which acts on frame tensors as
\begin{equation}
  \label{eq:general-affine-covariant-derivative}
  \D X^A{}_B = \der X^A{}_B + \Omega^A{}_C \wedge X^C{}_B - \Omega^C{}_B \wedge X^A{}_C.
\end{equation}
At this point, we take $\Omega^A{}_B$ to be a general (linear) affine connection, which takes values in the Lie algebra $\algebra{gl}{n,\RR}$.
The torsion two-form $T^A$ of the connection is given by
\begin{equation}
  \label{eq:affine-torsion}
  T^A = \der E^A + \Omega^A{}_B \wedge E^B.
\end{equation}
Additionally, under local $\group{GL}{n,\RR}$ transformations $\Lambda^A{}_B$,
the connection and a frame tensor $X^A{}_B$ transform as
\begin{align}
  \label{eq:affine-connection-transformation}
  \delta \Omega^A{}_B
  &= - \der \Lambda^A{}_B + \Lambda^A{}_C \Omega^C{}_B  - \Lambda^C{}_B \Omega^A{}_C,
  \\
  \delta X^A{}_B
  &= \Lambda^A{}_C X^C{}_B  - \Lambda^C{}_B X^A{}_C.
\end{align}
Finally, the frame bundle connection $\Omega^A{}_B$ can be related to an affine connection $\Gamma^\rho_{\mu\nu}$ acting on tensor products of the tangent and cotangent bundle by requiring the vielbeine to be parallel with respect to the sum of these connections,
\begin{equation}
  \label{eq:vielbein-postulate}
  0
  = \pd_\mu E^A_\nu - \Gamma^\rho_{\mu\nu} E^A_\rho + \Omega_\mu{}^A{}_B E_\nu^B.
\end{equation}
This relation is also known as the `vielbein postulate'.

So far, none of the above requires the existence of a metric.
Introducing a Minkowski metric~$\eta_{AB}$ on the frame bundle breaks the $\group{GL}{n,\RR}$ transformations of a general frame bundle down to the $\group{SO}{1,n-1}$ Lorentz transformations that leave it invariant.
The Lorentz algebra also corresponds to the local symmetries of the associated Lorentzian tangent bundle metric
\begin{equation}
  \label{eq:tangent-frame-metric-relation}
  g_{\mu\nu} = \eta_{AB} E^A_\mu E^B_\nu.
\end{equation}
The metric $\eta_{AB}$ may not be covariantly constant for a general affine connection, which we can measure using
\begin{equation}
  \label{eq;lorentzian-non-metricity}
  Q_{AB} = - \D \eta_{AB} / 2 = (\Omega_{AB} + \Omega_{BA})/2
  = \Omega_{(AB)}.
\end{equation}
We therefore refer to $Q_{AB}$ as the \emph{Lorentzian non-metricity tensor}.
Note that we have lowered one of the indices on the connection using $\eta_{AB}$, which allows us to
consider its (anti)symmetrization, so that we can split
\begin{gather}
  \label{eq:lorentzian-connection-split}
  \Omega_{AB} = \Omega_{[AB]} + Q_{AB},
  \\
  \label{eq:lorentzian-torsion-split}
  T^A = \der E^A + \Omega^{[AB]}\wedge E_B + Q^{AB} \wedge E_B.
\end{gather}
We can then use the torsion equation~\eqref{eq:lorentzian-torsion-split} to solve for $\Omega_{[AB]}$ in terms of the non-metricity~$Q_{AB}$ and torsion $T^A$,
\begin{equation}
  \label{eq:general-connection-decomposition}
  \Omega_{AB}
  = \hOmega_{AB}
  + K_{AB}
  + L_{AB}.
\end{equation}
This expresses a general $\algebra{gl}{n,\RR}$ connection $\Omega^A{}_B$
in terms of the Levi-Civita connection, which we denote by
\begin{equation}
  \label{eq:levi-civita-connection}
  \hOmega_{AB}
  = \der E_{[A}\left(\Theta_{B]},\Theta_C\right) E^C
  - \frac{1}{2} \der E_C\left(\Theta_A,\Theta_B\right) E^C,
\end{equation}
as well as the contorsion $K_{AB}$ and disformation $L_{AB}$,
\begin{align}
  K_{AB}(\Theta_C)
  &= \frac{1}{2} T_{C}(\Theta_A,\Theta_B)
  - T_{[A}(\Theta_{B]},\Theta_C),
  \\
  L_{AB}(\Theta_C)
  &= \frac{1}{2} Q_{AB}(\Theta_C)
  - Q_{C(A}(\Theta_{B)}).
\end{align}
Setting $T_A=0$ and $Q_{AB}=0$, the decomposition~\eqref{eq:general-connection-decomposition} implies the well-known fact that the Levi-Civita connection $\hOmega_{AB}$ is the unique torsion-free connection that preserves the Lorentzian metric $\eta_{AB}$.

Following Equation~\eqref{eq:lorentzian-connection-split}, we have thus seen that introducing a frame metric such as $\eta_{AB}$ naturally splits an affine connection $\Omega^A{}_B$ into a part $\Omega_{[AB]}$ that preserves the metric and a part $Q_{AB}=\Omega_{(AB)}$ that does not.
We will consider the Galilean version of this decomposition of a general $\algebra{gl}{n,\RR}$ connection below.

\paragraph{Palatini action and shift.}
We now turn to general relativity, where our starting point is the frame formulation of the Palatini action,
\begin{equation}
  \label{eq:standard-palatini-action}
  S_\text{Pal}[E,\Omega]
  = \frac{1}{2\kappa c} \int_M \eta_{AB} \ast (E^A \wedge E^C) \wedge R^B{}_C,
\end{equation}
with $\kappa=8\pi G_N c^{-4}$.
This action contains both the vielbeine $E^A$ and a $\algebra{gl}{n,\RR}$ connection $\Omega^A{}_B$ as variables.
The latter appears through the curvature two-form
\begin{equation}
  \label{eq:standard-curvature}
  R^A{}_B = \der \Omega^A{}_B + \Omega^A{}_C \wedge \Omega^C{}_B.
\end{equation}
Furthermore, the Hodge dual $\ast$ leads to a $(n-2)$-form
\begin{equation}
  \ast(E^A\wedge E^C)
  = \frac{\eta^{AD_1} \eta^{CD_2}}{(n-2)!} \epsilon_{D_1 \cdots D_n} E^{D_3} \wedge \cdots \wedge E^{D_n},
\end{equation}
so that the total integrand of the action~\eqref{eq:standard-palatini-action} is an $n$-form.
Here, $\epsilon_{A_1\cdots A_n}$ is the fully antisymmetric symbol.

Although the Palatini action is formulated using the Minkowski metric $\eta_{AB}$, we actually do not need to assume that the connection variable $\Omega^A{}_B$ is metric-compatible or torsionless.
Starting from a fully general $\algebra{gl}{n,\RR}$ connection $\Omega^A{}_B$, its equations of motion lead to
\begin{equation}
  \label{eq:general-connection-eom-solution}
  \Omega_{AB}
  = \hOmega_{AB} + \eta_{AB} Z.
\end{equation}
The one-form $Z$ is arbitrary but drops out of the action upon substituting the solution, and we can remove this ambiguity using a Lagrange multiplier for a constraint, see for example~\cite{Julia:1998ys,Dadhich:2012htv}.
Out of all possible $\algebra{gl}{n,\RR}$ connections,
the Palatini action then uniquely selects the Levi-Civita connection~\eqref{eq:levi-civita-connection}.
Substituting this solution in the action reproduces the Einstein--Hilbert action~\eqref{eq:standard-einstein-hilbert-action-lagrangian}.

Our goal is now to identify a first-order formulation of the PNR parametrization of the Einstein--Hilbert action~\eqref{eq:pnr-einstein-hilbert-lagrangian} in terms of the adapted connection $C^\rho_{\mu\nu}$ \eqref{eq:C-connection}.
This can easily be achieved using the linear field redefinition
\begin{equation}
  \label{eq:shifted-connection}
  \Omega^A{}_B = \tOmega^A{}_B + S^A{}_B,
\end{equation}
where the `shift' parameter $S^A{}_B$ is a particular function of the vielbeine and their derivatives.
The on-shell value of the new connection variable $\tOmega^A{}_B$ is then given by
\begin{equation}
  \label{eq:shifted-connection-on-shell}
  \left.\tOmega^A{}_B\right\vert_{\text{on-shell}}
    = \hOmega^A{}_B - S^A{}_B.
\end{equation}
As a result, even though the Palatini action~\eqref{eq:standard-palatini-action} implies that the on-shell value of $\Omega^A{}_B$ corresponds to the Levi-Civita connection,
an alternative connection can be obtained on shell from $\tOmega^A{}_B$ using a suitable shift $S^A{}_B$.

In the action~\eqref{eq:standard-palatini-action}, the field redefinition~\eqref{eq:shifted-connection} leads to
\begin{align}
  \label{eq:off-shell-shifted-action}
  S[E,\tOmega]
  = \frac{1}{2\kappa c} \int_M
  &
    \eta_{AB}\ast (E^A \wedge E^C)
    \vphantom{\left(\tilde{R}^B{}_C\right)}
    \\
  &{}\quad
    \wedge \left(
    \tilde{R}^B{}_C
      + S^B{}_D \wedge S^D{}_C
      + \tilde{\D} S^B{}_C
    \right).
  \nonumber
\end{align}
This `shifted' action now depends on $\tOmega^A{}_B$ and its associated covariant derivative $\tilde{D}$.
Through the field redefinition~\eqref{eq:shifted-connection}, it is equivalent to the Palatini action~\eqref{eq:standard-palatini-action}.
In the following, we will identify a suitable shift $S^A{}_B$ to obtain the frame equivalent of the desired PNR connection~\eqref{eq:C-connection}.
To show that the resulting action~\eqref{eq:off-shell-shifted-action} is equivalent to the PNR parametrization~\eqref{eq:pnr-einstein-hilbert-lagrangian} of the Einstein--Hilbert action, it is useful to rewrite it as
\begin{align}
  \label{eq:off-shell-shifted-action-rewritten}
  &S[E,\tOmega] =\nonumber\\
  &\quad \frac{1}{2\kappa c} \int_M \eta_{AB}\ast (E^A \wedge E^C) \wedge \left(
    \tilde R^B{}_C - S^B{}_D \wedge S^D{}_C
  \right)
  \nonumber\\
  &{}\qquad
  + \der \left[
    S^A{}_B \ast (E_A \wedge E^B)
  \right]
  + S^A{}_B \wedge \D \ast (E_A \wedge E^B)
  \nonumber\\
  &{}\qquad
  + \ast (E^A \wedge E^C) \wedge \left(
    \Omega_{AB} + \Omega_{BA}
  \right) \wedge S^B{}_C.
\end{align}
Here, we have introduced a boundary term and reinstated the covariant derivative $\D$ with respect to the original connection $\Omega^A{}_B$.
On shell, the latter is the Levi-Civita connection, which implies $\D E^A=0$ and $Q_{AB}=0$, so that integrating out the connection leads to
\begin{align}
  \label{eq:on-shell-shifted-action}
  S[E]
  = \frac{1}{2\kappa c} \int_M
  &\eta_{AB} \ast (E^A \wedge E^C)
  \\\nonumber
  &{}\qquad\quad
  \wedge \left[
    \widetilde{R}^B{}_C
    - S^B{}_D \wedge S^D{}_C
  \right].
\end{align}
Up to a boundary term, this action is equivalent to the Einstein--Hilbert Lagrangian of GR, for any choice of the shift parameter $S^A{}_B$.

\paragraph{Galilean connection and action.}
Just like how the Minkowski frame bundle metric $\eta_{AB}$ is associated to the Lorentzian tangent bundle metric $g_{\mu\nu}$ through Equation~\eqref{eq:tangent-frame-metric-relation}, we can associate two $\group{GL}{n,\RR}$ tensors
\begin{equation}
  \label{eq:galilean-invariant-tensors}
  t_A,
  \qquad
  \pi^{AB},
\end{equation}
satisfying $t_A \pi^{AB}=0$,
to the pre-non-relativistic (PNR) variables $T_\mu$ and $\Pi^{\mu\nu}$ defined in Equation~\eqref{eq:lorentzian-tangent-metric-pnr-parametrization}.
Whereas the Minkowski metric $\eta_{AB}$ is fixed by the Lorentz subalgebra of $\algebra{gl}{n,\RR}$, the tensors~\eqref{eq:galilean-invariant-tensors} are left invariant by the \emph{Galilean} algebra, which consists of spatial rotations and boosts with arbitrary velocity.
We also introduce the dual tensors $t^A$ and $\pi_{AB}$ such that we have the orthogonality and completeness relations
\begin{subequations}
  \begin{gather}
    \label{eq:galilean-orthogonality}
    t_A t^A = -1,
    \quad
    t^A \pi_{AB} = t_A \pi^{AB} = 0,
    \\
    \label{eq:galilean-completeness-relations}
    \delta^A_B = - t^A t_B + \pi^{AC}\pi_{CB},
  \end{gather}
\end{subequations}
corresponding to the same relations for the metric variables in Equation~\eqref{eq:orthogonality-completeness-relations-metric}.

However,
following the `pre-non-relativistic' decomposition~\eqref{eq:lorentzian-tangent-metric-pnr-parametrization} of the Lorentzian metric $g_{\mu\nu}=\eta_{AB}E^A_\mu E^B_\nu$,
we can decompose the Minkowski metric as
\begin{equation}
  \eta_{AB}
  = - t_A t_B + \pi_{AB},
  \quad
  \eta^{AB}
  = - t^A t^B + \pi^{AB},
\end{equation}
where we identify
\begin{subequations}
  \begin{alignat}{2}
    \label{eq:pnr-vielbein-variables}
    t_A E^A_\mu
    &= c\, T_\mu
    \qquad&
    \pi_{AB} \Theta^A_\mu \Theta^B_\mu
    &= \Pi_{\mu\nu},
    \\
    t^A \Theta_A^\mu
    &= - V^\mu /c,
    \qquad&
    \pi^{AB} \Theta_A^\mu \Theta_B^\nu
    &= \Pi^{\mu\nu}.
  \end{alignat}
\end{subequations}
The frame tensors $t_A$ and $\pi^{AB}$ (and hence also $T_\mu$ and $\Pi^{\mu\nu}$) can be shown to be invariant under local Galilean transformations, but their duals $t^A$ and $\pi_{AB}$ (and hence also $V^\mu$ and $\Pi_{\mu\nu}$) are not invariant.

We now want to recast the Palatini action~\eqref{eq:standard-palatini-action} in terms of the PNR Galilean variables defined above.
Before expanding in powers of $1/c^2$, the action and the frames are invariant under the Lorentz algebra, and this Galilean description is somewhat unnatural.
However, as outlined in the Introduction, by rewriting the action for GR in this way, we are \emph{preparing} ourselves for a covariant description of the non-relativistic Newton--Cartan geometry and its Galilei symmetry that appears in the expansion.

Our main task now is to introduce a suitable Galilean connection, corresponding to the frame version of the $C^\rho_{\mu\nu}$ connection~\eqref{eq:C-connection} that was mentioned in the Introduction, and to identify its associated shift $S^A{}_B$.
The resulting connection is compatible with the Galilean metric structure~\eqref{eq:galilean-invariant-tensors} and has only the minimal amount of torsion required for a Galilean connection, as detailed below.

First, in analogy with the Lorentzian definition~\eqref{eq;lorentzian-non-metricity},  we define the Galilean non-metricities as
\begin{subequations}
  \label{eq:galilean-metricity}
  \begin{gather}
    \gQ_A
    = - \D t_A
    = t_B \Omega^B{}_A,
    \\
    \gQ^{AB} = \D \pi^{AB} /2
    = (\pi^{AC} \Omega^B{}_C + \pi^{CB}\Omega^A{}_C)/2.
  \end{gather}
\end{subequations}
It is useful to split the tangent space index $A=(0,a)$ into space and time components, so that we can write the Galilean tensors~\eqref{eq:galilean-invariant-tensors} as
\begin{gather}
  \label{eq:galilean-invariant-tensors-adapted-coords}
  t_A = \delta_A^0,
  \quad
  \pi^{AB} = \delta^A_a \delta^B_b \delta^{ab},
  \\
  \label{eq:gauge-fixed-basis}
  E^A = (cT, \mE^a),
  \quad
  \Theta^A = (- V/c, \Theta^a).
\end{gather}
The non-zero components of the non-metricities~\eqref{eq:galilean-metricity} are
\begin{align}
  \label{eq:galilean-non-metricity-def}
  \ost{Q}{G} = \Omega^0{}_0\,,
  \quad
 \ost{Q}{G}_a =  c\inv \Omega^0{}_a\,,
  \quad
  \ost{Q}{G}^{ab} = \Omega^{(ab)}\,.
\end{align}
Here and in the following, we have introduced the appropriate factors of $c$ so that the resulting objects such as $\gQ_a$ are $\order{c^0}$.
Any connection for which all these components vanish is compatible with the Galilean PNR structure, and will therefore lead to a Newton--Cartan metric-compatible connection upon expansion.

Note that we can raise and lower spatial indices using  $\delta_{ab}$ and $\delta^{ab}$.
Together with~\eqref{eq:galilean-non-metricity-def}, this allows us to split a general $\algebra{gl}{n,\RR}$ connection into
\begin{equation}
  \label{eq:galilean-connection-split}
  \Omega^{[ab]},
  \quad
  \Omega^a{}_0,
  \quad
  \gQ^{ab},
  \quad
  \gQ,
  \quad
  \gQ_a,
\end{equation}
in analogy with the Lorentzian decomposition~\eqref{eq:lorentzian-connection-split}.
The torsion equations~\eqref{eq:affine-torsion} then take the form
\begin{align}
  \label{eq:torsion-equations-galilean}
  T^0
  &= c \der T + c \gQ_a \wedge \mE^a + c \gQ \wedge T,
  \\ \nonumber
  T^a
  &= \der \mE^a + \Omega^{[ab]} \wedge \mE_b
  + c \Omega^a{}_0 \wedge T
  + \gQ^{ab} \wedge \mE_b.
\end{align}
For Galilean connections with vanishing non-metricity, the time component of the torsion $T^0=c\der T$ is fixed independent of the remaining connection components~\footnote{%
  See Ref.~\cite{Figueroa-OFarrill:2020gpr} for a discussion on intrinsic torsion of non-relativistic geometries in the language of G-structures.
}.
This torsion is generically non-zero, and requiring that it vanishes would mean that the $1/c^2$ expansion of the geometry does not accommodate time dilation.

Instead, we allow the time component of the torsion $T^0=c\der T$ to be nonzero and only require that the spatial torsion $T^a$ vanishes.
As is known from a tangent bundle perspective~\cite{Hartong:2015zia}, this combination of torsion constraints and Galilean metric compatibility does not fully fix the connection.
Here, we choose to use the connection
\begin{subequations}
  \label{eq:omega-c-connection}
  \begin{align}
    \gOmega_{[ab]}
    &= \der \mE_{[a}\left(\Theta_{b]},\Theta_c\right)\mE^c
    - \der \mE_{[a}(\Theta_{b]},V) T
    \\\nonumber
    &{}\qquad
    - \frac{1}{2} \der \mE_c\left(\Theta_a,\Theta_b\right) \mE^c
    \\
    \gOmega^{a}= c\Omega^a{}_0
    &= \frac{1}{2} \left[
     \der  \mE^a(\Theta_c,V) + \der  \mE_c(\Theta^a,V)
    \right] \mE^c,
  \end{align}
\end{subequations}
This is the frame equivalent of the $C^\rho_{\mu\nu}$ connection~\eqref{eq:C-connection} that naturally appears in the pre-non-relativistic decomposition of the Levi-Civita connection~\cite{Hansen:2020pqs}, as can be seen using the vielbein postulate~\eqref{eq:vielbein-postulate}.
From an algebraic perspective, the connections~\eqref{eq:omega-c-connection} correspond to spatial rotations and Galilean boosts, see the appendix.

The Galilean connection~\eqref{eq:omega-c-connection} can be obtained as the on-shell value~\eqref{eq:shifted-connection-on-shell} for $\tilde{\Omega}^A{}_B$,
using the shift
\begin{subequations}
  \label{eq:c-shift}
  \begin{align}
    S^a{}_b
    &= \frac{c^2}{2} \der T(\Theta^a,\Theta_b) T,
    \\
    S^a{}_0
    &= \frac{c}{2} \der T(\Theta^a,\Theta_c) \mE^c
    - c \der T(\Theta^a, V) T,
    \\
    S^0{}_b
    &= \frac{c}{2} \der T(\Theta_b,\Theta_c)\mE^c
    - c \der T(\Theta_b,V) T
    \\\nonumber
    &{}\qquad
    + \frac{1}{2c} [
     \der \mE_b(\Theta_c,V) + \der \mE_c(\Theta_b,V)
    ] \mE^c.
  \end{align}
\end{subequations}

We can now work out the terms in the second-order action~\eqref{eq:on-shell-shifted-action} explicitly in terms of Galilean-covariant objects.
First, we have
\begin{align}
  & -\eta_{AB} \ast(E^A \wedge E^C) \wedge S^B{}_D \wedge S^D{}_C
  \nonumber\\
  &{}\qquad
  = \delta^{ab} \delta^{cd} \frac{c^2}{4} \der T(\Theta_a,\Theta_c) \der T(\Theta_b,\Theta_d),
\end{align}
which is equivalent to the leading term in the Lagrangian~\eqref{eq:pnr-einstein-hilbert-lagrangian}.
The connection~\eqref{eq:omega-c-connection} is metric-compatible, which means that all non-metricities~\eqref{eq:galilean-non-metricity-def} vanish.
As a result, the components of its curvature~\eqref{eq:standard-curvature} are given by
\begin{subequations}
  \begin{align}
    R^a{}_b
    &= d\Omega^a{}_b
    + \Omega^a{}_c \wedge \Omega^c{}_b
    \\
    R^a
    = c R^a{}_0
    &= d\Omega^a + \Omega^a{}_b \wedge \Omega^b.
  \end{align}
\end{subequations}
The combination of these curvatures that enters in the action is invariant under local Lorentz transformations before expanding in powers of $1/c^2$,
and after expanding it is invariant under local Galilean boosts at each order.

We can now write the action with on-shell connection~\eqref{eq:on-shell-shifted-action} in a Galilean-covariant way as
\begin{align}
  \label{eq:on-shell-c-shifted-action}
  S
  = \frac{1}{2\kappa } \int_M
  &\left[
    \frac{c^2}{4} \der T(\Theta_a,\Theta_b) \der T(\Theta^a,\Theta^b)
  \right.
  \\\nonumber
  &{}\quad\left.
  + R^a{}_b (\Theta_a,\Theta^b)
  - \frac{1}{c^2} R^a(V,\Theta_a)
  \right]
  E  \der^{n} x.
\end{align}
with $E=\det(T_\mu,\mE^a_\mu)$.
This action is still Lorentz invariant, but it is constructed using Galilean building blocks.
It precisely reproduces the PNR form of the Einstein--Hilbert Lagrangian~\eqref{eq:pnr-einstein-hilbert-lagrangian}.
Given the shift~\eqref{eq:c-shift}, the equivalent decomposition of the off-shell action~\eqref{eq:off-shell-shifted-action-rewritten} in terms of a Galilean-compatible covariant derivative and non-metricity can then be used for the non-relativistic expansion of general relativity in the first-order formulation.

\paragraph{Discussion and outlook.}
Through an appropriate field redefinition, we have obtained an alternative formulation of the Palatini action,
whose equations of motion
give rise to a connection
that is adapted to the Galilean-covariant Newton--Cartan structure that arises in a non-relativistic $1/c^2$ expansion of GR.
By identifying the Lorentzian and Galilean non-metricity associated to a general affine connection, we also see how our field redefinition naturally leads to nonzero torsion.
After substituting the connection, the resulting action is equivalent to GR.
As such, we can reproduce the geometric, off-shell non-relativistic expansion of GR that was recently developed in a second-order formulation~\cite{Hansen:2019pkl,Hansen:2020pqs}, and our first-order approach will allow it to be extended to higher orders in a more tractable manner. Importantly, our first-order description is expected
to be essential in establishing the connection between the $1/c^2$~expansion of GR and the PN approximation (as shown to 1PN order in \cite{Tichy:2011te}).
Furthermore, the expansion can also be applied to a wide range problems in astrophysics where velocities are low.
We also expect that this action will be of use in computing conserved boundary charges, since this is typically more accessible in a first-order formulation.

Additionally, we can apply our procedure to the ultra-relativistic $c\to0$ limit, where the Galilean local symmetry is replaced by the Carroll algebra~\cite{LeBellac,Bergshoeff:2017btm}.
The resulting expansion in powers of $c^2$ gives a novel approach to Carroll geometry (together with its subleading corrections), which is connected to the geometry of null surfaces in GR~\cite{Hartong:2015xda} and is therefore potentially relevant to the dynamics of black hole horizons~\cite{Donnay:2019jiz} and the holographic description of asymptotically flat space at null infinity~\cite{Duval:2014uva,Ciambelli:2019lap}.
This geometry is constructed using the Carroll-invariant tensors $t^A$ and $\pi_{AB}$, which have opposite index structure compared to their Galilei counterparts.
In this sense, the Carrollian notion of metric-compatibility is the dual of the Galilean case considered above, see also Ref.~\cite{Duval:2014uoa}.
We will address the ultra-relativistic expansion of our novel action and the corresponding geometry in more detail in upcoming work~\cite{Hansen:2021b}.
Different approaches to frame and/or first-order formulations of non-relativistic and ultra-relativistic limits and expansions of gravity have previously been considered in Refs.~\cite{Bergshoeff:2017btm,Cariglia:2018hyr,Guerrieri:2020vhp}.

Finally, our field redefinition procedure can also be used to obtain Palatini-type actions for GR whose equation of motion lead to a connection with vanishing Riemann tensor but non-trivial torsion or non-metricity.
This corresponds to the teleparallel or symmetric teleparallel formulation of GR, respectively, see also~\cite{BeltranJimenez:2018vdo}.

\paragraph{Acknowledgments.}
\begin{acknowledgments}
We thank Stefano Baiguera, Jørgen Sandøe Musaeus, Benjamin Tangen Søgaard and especially Dieter Van den Bleeken and Manus Visser for useful discussions.
The work of DH is supported by the Swiss National Science Foundation through the NCCR SwissMAP.
The work of JH is supported by the Royal Society University Research Fellowship ``Non-Lorentzian Geometry in Holography'' (grant number UF160197).
The work of NO and GO is supported in part by the project ``Towards a deeper understanding of  black holes with non-relativistic holography'' of the Independent Research Fund Denmark (grant number DFF-6108-00340) and the Villum Foundation Experiment project 00023086.
\end{acknowledgments}

\section{Appendix: algebraic perspective}
We have considered Lorentzian and Galilean decompositions of a general affine $\algebra{gl}{n,\RR}$ connection based on the corresponding notions of metric-compatibility in the main text.
Alternatively, we can understand our results from a more algebraic perspective.
One can obtain the geometry associated to a generic frame bundle from a gauge connection $\mA$ that is associated to the affine group $\RR^n \rtimes \group{GL}{n,\RR}$, which we can parametrize as
\begin{equation}
  \label{eq:app-general-affine-connection}
  \mA = E^A P_A + \frac{1}{2} \Omega^A{}_B \, M_A{}^B.
\end{equation}
Here, the translations $P_A$ and the matrices $M_A{}^B$ generate the affine algebra,
\begin{subequations}
  \begin{align}
    [M_A{}^B, M_C{}^D]
    &= - \delta_A^D M_C{}^B + \delta_C^B M_A{}^D
    \\
    [M_A{}^B, P_C]
    &= \delta_C^B P_A.
  \end{align}
\end{subequations}
Given a Minkowski metric~$\eta_{AB}$ we can raise and lower indices on
the matrix generators $M_{AB}=\eta_{BC}M_A{}^C$
and the connection $\Omega^{AB} = \eta^{BC} \Omega^A{}_C$,
so that we can split them in symmetric and antisymmetric components,
\begin{subequations}
  \begin{gather}
  S_{AB} = M_{(AB)},
  \quad
  J_{AB} = M_{[AB]},
  \\
  \mA = E^A P_A
  + \frac{1}{2} \Omega^{[AB]} \, J_{AB}
  + \frac{1}{2} Q^{AB} \, S_{AB}.
  \end{gather}
\end{subequations}
The antisymmetric $J_{AB}$ generate the Lorentz algebra and couple to the metric-compatible connection~$\Omega^{[AB]}$, whereas the symmetric generators $S_{AB}$ couple to the Lorentzian non-metricity $Q^{AB} = \Omega^{(AB)}$.
This corresponds to the decomposition in Equation~\eqref{eq:lorentzian-connection-split}.

Alternatively, using $A=(0,a)$ corresponding to Equation~\eqref{eq:galilean-invariant-tensors-adapted-coords}, we can decompose the affine algebra as
\begin{equation}
  \begin{gathered}
    J_{ab} = M_{[ab]},
    \quad
    G_a = M_a{}^0,
    \quad
    C^a = M_0{}^a\,,
    \\
    H = P_0,
    \quad
    P_a,
    \quad
    S_{ab} = M_{(ab)},
    \quad
    S_0{}^0 = M_0{}^0\,.
  \end{gathered}
\end{equation}
Here, the translations are split in time translations~$H$ and space translations $P_a$, and the spatial rotations $J_{ab}$ and Galilean boosts $G_a$ generate the Galilei subalgebra of $\algebra{gl}{n,\RR}$.
This induces the decomposition of the affine connection~\eqref{eq:app-general-affine-connection}
in the Galilean-compatible components
$\Omega^{[ab]}$
and
$\Omega^a = c\Omega^a{}_0$ as well as the non-metricities
\begin{equation}
  \gQ^{ab} = \Omega^{(ab)},
  \quad
  \gQ_a = c\inv \Omega^0{}_a,
  \quad
  \gQ = \Omega^0{}_0,
\end{equation}
corresponding to Equation~\eqref{eq:galilean-connection-split}.

Finally, note that $J_{ab}$ and $C^a$ generate the Carroll subalgebra of $\algebra{gl}{n,\RR}$, which induces the corresponding Carrollian decomposition of $\Omega^A{}_B$ that was briefly mentioned in the Conclusion.
In particular, we see that the Galilean boost generator $G_a$ corresponds to a Carrollian non-metricity component, while the Carrollian boost generator $C_a$ corresponds to a Galilean non-metricity component.
In this sense, the Carroll decomposition of a general affine connection is therefore the dual of the Galilean decomposition, see also Ref.~\cite{Duval:2014uoa}.

\bibliographystyle{apsrev4-1}
\bibliography{letter}

\end{document}